\documentclass[journal,twoside]{IEEEtran}
\usepackage[dvipdfmx]{graphicx}
\usepackage{amsmath}
\usepackage{bm}
\usepackage{here}
\usepackage{scalefnt}
\usepackage{cite}
\usepackage{nidanfloat}
\usepackage{mathtools}
\usepackage{amssymb}
\usepackage{url}
\usepackage{color}
\definecolor{MYBLUE}{rgb}{0,0,1}

\def\qed{\relax\ifmmode\hskip2em \Box\else\unskip\nobreak\hskip1em $\Box$\fi}
\begin{document}
\title{An Adaptive Periodic-Disturbance Observer\\for Periodic-Disturbance Suppression}
\author{
Hisayoshi~Muramatsu,~\IEEEmembership{Student~Member,~IEEE,}
and~Seiichiro~Katsura,~\IEEEmembership{Member,~IEEE}
\thanks{
	\color{blue}
	\copyright 2018 IEEE.  Personal use of this material is permitted.  Permission from IEEE must be obtained for all other uses, in any current or future media, including reprinting/republishing this material for advertising or promotional purposes, creating new collective works, for resale or redistribution to servers or lists, or reuse of any copyrighted component of this work in other works.\\\indent
	Hisayoshi Muramatsu and Seiichiro Katsura, “An Adaptive Periodic-Disturbance Observer for Periodic-Disturbance Suppression,” IEEE Transactions on Industrial Informatics, vol. 14, no. 10, pp. 4446–4456, Oct. 2018.\\\noindent
	Digital Object Identifier 10.1109/TII.2018.2804338\\\indent
	The published version of the paper is available online at https://doi.org/10.1109/TII.2018.2804338
	\color{blue}
}
\thanks{H. Muramatsu and S. Katsura are with the Department of System Design Engineering, Keio University, Yokohama 223-8522, Japan (e-mail: muramatsu@katsura.sd.keio.ac.jp; katsura@sd.keio.ac.jp).}
}
\markboth{{\color{myblue}The published version of the paper is available online at https://doi.org/10.1109/TII.2018.2804338}}{{\color{myblue}The published version of the paper is available online at https://doi.org/10.1109/TII.2018.2804338}}

\maketitle
\begin{abstract}
Repetitive operations are widely conducted by automatic machines in industry.
Periodic disturbances induced by the repetitive operations must be compensated to achieve precise functioning.
In this paper, a periodic-disturbance observer (PDOB) based on the disturbance observer (DOB) structure is proposed.
The PDOB compensates a periodic disturbance including the fundamental wave and harmonics by using a time delay element.
Furthermore, an adaptive PDOB is proposed for the compensation of frequency-varying periodic disturbances.
An adaptive notch filter (ANF) is used in the adaptive PDOB to estimate the fundamental frequency of the periodic disturbance.
Simulations compare the proposed methods with a repetitive controller (RC) and the DOB.
Practical performances are validated in experiments using a multi-axis manipulator.
The proposal provides a new framework based on the DOB structure to design controllers using a time delay element.
\end{abstract}

\begin{IEEEkeywords}
Periodic disturbance, disturbance observer, adaptive notch filter, time delay element, adaptive control
\end{IEEEkeywords}

\IEEEpeerreviewmaketitle
\section{Introduction}
\IEEEPARstart{P}{eriodic} motions are usual industrial tasks in operations using automatic systems.
For example, actuators, automatic machines, and robots are used because of their high precision, high speed, and their capability to work for long hours.
However, most repetitive activities induce periodic disturbances that consist of a fundamental wave and harmonics.
In order to realize precise periodic motions, periodic-disturbance suppression considering the fundamental wave and harmonics is a problem to be solved.

A disturbance observer (DOB) is a two-degree-of-freedom controller based on an observer structure for suppressing disturbances \cite{DOBoverview,DOBohnishi,ObBasedPeriodic,NatoriCDOB}.
The two-degree-of-freedom structure can only control the disturbance suppression to not affect the tracking ability.
The DOB has a Q-filter $Q(z^{-1})$ that determines the sensitivity function $Q(z^{-1})z^{-1}$ and the complementary sensitivity function $1-Q(z^{-1})z^{-1}$of the DOB.
The Q-filter is typically set to a low-pass filter to realize high-pass sensitivity and low-pass complimentary sensitivity on the basis of a tradeoff between the functions because the sensitivity characteristic enables disturbance suppression and the complementary sensitivity characteristic corresponds to noise sensitivity and robust stability \cite{DOBAna,DOBHinf}.
However, the high-pass characteristic is not suitable to compensate a periodic disturbance because it possesses powers only at an infinite number of specific frequencies.
To suppress the periodic disturbance, DOB-based controllers and high-order DOBs have been studied by focusing on some specific high-frequency waves \cite{CH2,DOBsinD}.
In \cite{DOBperiod}, a periodic adaptive disturbance observer was proposed that consists of searching and learning phases and considers harmonics.
A typical DOB is only used in the searching phase as an initial condition for the learning phase.
Hence, it does not use the advantages of a DOB structure in the learning phase to compensate disturbances.
Since the method (which also affects the tracking characteristic) is not the two-degree-of-freedom controller, it is more similar to a repetitive controller (RC) rather than to a DOB.

Repetitive control  is a well-known method that focuses on suppressing all frequencies of a periodic disturbance \cite{RC1,RCPWM,RPTCFujimoto2}.
The RC improves both tracking performance regarding a periodic signal and suppression performance on an exogenous periodic signal by using a time delay element.
However, RCs have the following three problems.
First, the time delay element interferes with the command-tracking characteristic and the nominal stability.
Second, the sensitivity characteristic amplifies other disturbances such as aperiodic disturbances.
Finally, the complementary sensitivity function is not sufficiently designed.

RCs using a DOB structure have been studied \cite{DOBbasedRLC}.
In \cite{RepetitiveDOB1}, attenuating the amplification of other disturbances was achieved.
However, the bandwidth of the band-stop characteristics suppressing periodic disturbances is narrow, and the command-tracking characteristic and the nominal stability are still affected by the time delay element.

In this paper, a periodic disturbance observer (PDOB) is proposed.
It contributes a simultaneous realization of the characteristics, which solves the above-mentioned problems.
\begin{enumerate}
	\item The PDOB including a time delay element aims at compensating all frequencies of a periodic disturbance with an infinite number of band-stop characteristics.
	\item The time delay element does not affect the nominal stability.
	\item The PDOB does not affect the nominal command-tracking characteristic.
	\item The amplification of other disturbances can be attenuated by the sensitivity function adjusted by a design parameter $\gamma$.
	\item The parameter $\gamma$ can also adjust the complementary sensitivity function.
	\item Implementation is simple because of using the time delay element.
\end{enumerate}
The PDOB achieves performance by focusing only on periodic-disturbance suppression.
Thus, it can estimate and compensate periodic disturbances more effectively than conventional DOBs because the PDOB uses a time delay element considering the dynamics of a periodic disturbance.
Further, owing to the integration of the time delay element, the PDOB leads better results regarding periodic disturbances than other conventional disturbance estimation methods, e.g., active disturbance rejection control and an uncertainty-and-disturbance estimator \cite{DRC,UDE}.

Moreover, deterioration of the suppression performance due to, e.g., identification errors, unknown frequencies, or varying frequencies is also an important problem \cite{UnPdis,AdaptiveNarrowDisBenchmark}.
To solve problems due to uncertainties, adaptive control is a typical approach \cite{CooperativeAdaptiveCrane,AdaptiveNNuncertainRobot}.
This paper specifically addresses adaptive estimation of a fundamental frequency regarding a periodic disturbance.
In recent years, adaptive rejection of several sinusoidal disturbances \cite{AdaptiveNarrowDisReview} and RCs based on a multiple-memory-loop technique broadening the bandwidth \cite{Steinbuch20022103} have been studied.
This paper proposes an adaptive PDOB in addition to the PDOB to compensate also frequency-varying periodic disturbances.
Since adaptive notch filters (ANFs) are well-known frequency estimators regarding periodic signals \cite{AdaptiveNotch1,ANFsurvey}, the adaptive PDOB employs an ANF that is designed to estimate a fundamental frequency from a periodic disturbance including harmonics.
The adaptive PDOB acquires adaptivity through the design of six additional design parameters.

\section{Periodic-Disturbance Observer} \label{sec2}
\subsection{Q-filter and Transfer Functions} \label{sec2-1}
\begin{figure*}[t]
	\begin{minipage}{0.33\hsize}
		\begin{center}
		\includegraphics[width=\hsize]{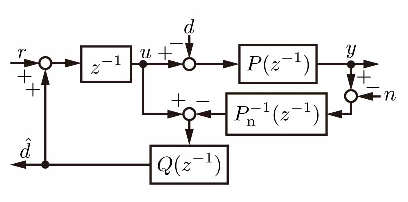}
		\end{center}
	\end{minipage}
	\begin{minipage}{0.33\hsize}
		\begin{center}
		\includegraphics[width=\hsize]{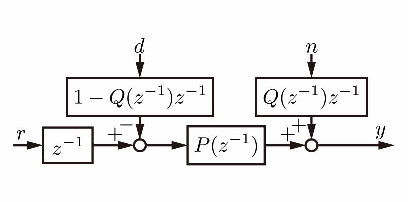}
		\end{center}
	\end{minipage}
	\begin{minipage}{0.33\hsize}
		\begin{center}
		\includegraphics[width=\hsize]{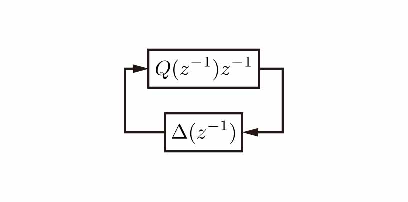}
		\end{center}
	\end{minipage}\\
	\begin{minipage}{0.33\hsize}
		\begin{center}
		(a)
		\end{center}
	\end{minipage}
	\begin{minipage}{0.33\hsize}
		\begin{center}
		(b)
		\end{center}
	\end{minipage}
	\begin{minipage}{0.33\hsize}
		\begin{center}
		(c)
		\end{center}
	\end{minipage}
	\caption{(a) Block diagram of general DOB. (b) Equivalent Block diagram of DOB for $\Delta(z^{-1})=0$. (c) Equivalent Block diagram of DOB for small-gain theorem.}
	\label{fig:DOBs}
\end{figure*}

A disturbance $d(k)$ is defined as a composition of a periodic disturbance $d_{\mathrm{p}}(k)$ and an aperiodic disturbance $d_{\mathrm{a}}(k)$
\begin{align}
	d(k)=d_{\mathrm{p}}(k)+d_{\mathrm{a}}(k).\notag
\end{align}
The dynamics of the periodic disturbance are defined by
\begin{align}
	d_{\mathrm{p}}(k)&=d_{\mathrm{p}}(k-N)+\rho(k),\notag
\end{align}
where
\begin{align}
	\rho(k)=\left\{
		\begin{array}{ll}
			\rho_0(k)&k<N\\
			0&N\leq k
		\end{array}
	\right. .
\end{align}
$\rho_0(k)$ and $N$ are the initial wave and delay, respectively.
The $Z$-transformed periodic disturbance is
\begin{align}
	d_{\mathrm{p}}(z^{-1})=\frac{1}{1-z^{-N}}\rho(z^{-1}).\notag
\end{align}
Fig.~\ref{fig:DOBs} shows a block diagram and its equivalents of a general DOB.
$r$, $n$, $y$, $u$, $\hat{\mbox{ }}$, $P$, $P_{\mathrm{n}}$, and $\Delta$ denote the reference, noise, output, input, estimated value, plant, nominal plant, and modeling error, respectively.
Regarding Fig.~\ref{fig:DOBs}(b), the periodic disturbance is compensated by the DOB with $\frac{1-Q(z^{-1})z^{-1}}{1-z^{-N}}\rho(z^{-1})$.
Since $\rho(k)$ is zero for $N\leq k$, the Q-filter satisfying
\begin{align}
	\frac{1-Q(z^{-1})z^{-1}}{1-z^{-N}}\rho(z^{-1})=\gamma\rho(z^{-1})\notag
\end{align}
suppresses the periodic disturbance for $N\leq k$.
The design parameter $\gamma$ is an integer.
The Q-filter of the PDOB is calculated as
\begin{align}
	\label{eq:PDOB:Q}
	Q(z^{-1})=q(z^{-1})\{1-\gamma(1-z^{-N})\},
\end{align}
where a low-pass filter $q(z^{-1})$ is added to improve robust stability and the $Z$-operator $z^{-1}$ is ignored to be a causal filter.
In this paper, the low-pass filter $q(z^{-1})$ is set to a first-order low-pass filter with
\begin{align}
	\label{eq:PDOB:qLPF}
	q(e^{-j\omega T_{\mathrm{s}}})=\frac{g}{g+j\omega},
\end{align}
where $g$ denotes the cutoff frequency.
The PDOB implementation requires the inverse-nominal-plant model $P^{-1}_{\mathrm{n}}$ and the three parameters: delay $N$, design parameter $\gamma$, and cutoff frequency $g$.
\begin{figure*}[b]
\hrulefill
\begin{align}
	\label{eq:TF_PDOB}
	\left[
		\begin{array}{c}
			y\\
			u\\
			\hat{d}
		\end{array}
	\right]&=
	\frac{1}{\phi}
	\left[
		\begin{array}{ccc}
			N_P^2D_Q(1+\Delta )z^{-1}&
			-N_P^2(D_Q-N_Qz^{-1})(1+\Delta )&
			N_PN_QD_P(1+\Delta )z^{-1}\\
			N_PD_PD_Qz^{-1}&
			N_PN_QD_P(1+\Delta )z^{-1}&
			N_QD_P^2z^{-1}\\
			-N_PN_Q\Delta D_Pz^{-1}&
			N_PN_QD_P(1+\Delta )&
			N_QD_P^2\\
		\end{array}
	\right]
	\left[
		\begin{array}{c}
			r\\
			d\\
			n
		\end{array}
	\right]
\end{align}
\end{figure*}

\begin{figure}[t]
	\begin{center}
		\includegraphics[width=0.9\hsize]{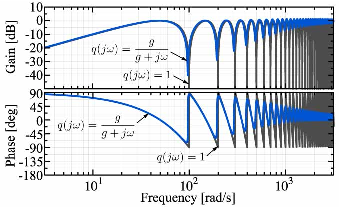}\\
		(a)\\
		\includegraphics[width=0.9\hsize]{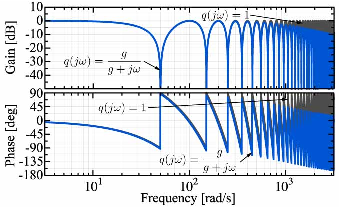}\\
		(b)
	\end{center}
	\caption{Effect of low-pass filter $q(z^{-1})$ on sensitivity and complementary sensitivity functions. The parameters are $N=2\pi/(\omega_0T_{\mathrm{k}})$, $\omega_0=100\ \mathrm{rad/s}$, $\gamma=0.5$, $g=1000\ \mathrm{rad/s}$, and $T_{\mathrm{k}}=0.01\ \mathrm{ms}$. (a) Sensitivity functions of PDOB, $1-Q(z^{-1})z^{-1}$. (b) Complementary sensitivity functions of PDOB, $Q(z^{-1})z^{-1}$.}
		\label{fig:LPF}
\end{figure}

The effects of the low-pass filter $q(z^{-1})$ on the sensitivity and complementary sensitivity functions are shown in Fig.~\ref{fig:LPF}.
The periodic-disturbance suppression and the robust stability are a tradeoff of both functions.
The low-pass filter is added to improve the complementary sensitivity function in the high-frequency range, as shown in Fig.~\ref{fig:LPF}(b).
The band-stop characteristics attenuated by the tradeoff are sufficient to compensate a periodic disturbance, as shown in Fig.~\ref{fig:LPF}(a).
It is because most real harmonics become weak in the high-frequency range.
The application of a high-order low-pass filter could improve the performance of the PDOB.
However, this study employs a first-order low-pass filter to demonstrate fundamental characteristics of the proposed methods.

The nine transfer functions regarding Fig.~\ref{fig:DOBs}(a) with $r$, $d$, and $n$ to $y$, $u$, and $\hat{d}$ are displayed in \eqref{eq:TF_PDOB}.
In the functions, the nominal plant $P_{\mathrm{n}}(z^{-1})$ and the Q-filter $Q(z^{-1})$ are expressed by the ratios of the coprime polynomials $\frac{N_P(z^{-1})}{D_P(z^{-1})}$ and $\frac{N_Q(z^{-1})}{D_Q(z^{-1})}$, respectively.
The modeling error $\Delta(z^{-1})$ is defined by
\begin{align}
	P(z^{-1})=\{1+\Delta(z^{-1})\}P_{\mathrm{n}}(z^{-1}).\notag
\end{align}
The characteristic equation is
\begin{align}
	\label{eq:PDOB:ChEq}
	\phi(z^{-1})&=N_PD_P(D_Q+N_Q\Delta z^{-1}).
\end{align}

\subsection{Parameter Design} \label{sec2-2}
\subsubsection{Nominal Stability}
Under a nominal condition, $\Delta(z^{-1})=0$, the characteristic equation in \eqref{eq:PDOB:ChEq} becomes
\begin{align}
	\phi_{\mathrm{n}}(z^{-1})&=N_PD_PD_Q.\notag
\end{align}
Since the polynomials of the Q-filter are $N_Q(z^{-1})=N_q(z^{-1})\{1-\gamma(1-z^{-N})\}$ and $D_Q(z^{-1})=D_q(z^{-1})$ with the ratio of the coprime polynomials of the low-pass filter, $q(z^{-1})=\frac{N_q(z^{-1})}{D_q(z^{-1})}$, the nominal stability of the PDOB depends only on the stabilities of the three elements: zeros of nominal plant $N_P$, poles of nominal plant $D_P$, and poles of the low-pass filter $D_Q$.
As a major characteristic of the PDOB, the nominal stability is independent of the delay $z^{-N}$.
\begin{figure}[t]
	\begin{center}
		\includegraphics[width=0.9\hsize]{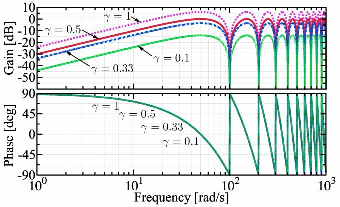}\\
		(a)\\
		\includegraphics[width=0.9\hsize]{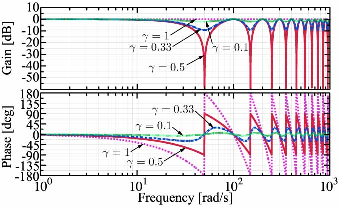}\\
		(b)
	\end{center}
		\caption{Sensitivity and complementary sensitivity functions with variations in $\gamma$. The parameters are $N=2\pi/(\omega_0T_{\mathrm{k}})$, $\omega_0=100\ \mathrm{rad/s}$, $q(z^{-1})=1$, and $T_{\mathrm{k}}=0.01\ \mathrm{ms}$. (a) Sensitivity functions of PDOB, $1-Q(z^{-1})z^{-1}$. (b) Complementary sensitivity functions of PDOB, $Q(z^{-1})z^{-1}$.}
		\label{fig:gamma}
\end{figure}
\subsubsection{Calculation of $N$}
The delay can be calculated with $N=\frac{2\pi}{T_{\mathrm{k}}\omega_0}$ considering the relation between period and angular frequency.
$T_{\mathrm{k}}$ and $\omega_0$ denote the sampling time and fundamental frequency, respectively.
However, the added low-pass filter $q(z^{-1})$ shifts the band-stop frequencies of $1-Q(z^{-1})z^{-1}$, as shown in Fig.~\ref{fig:LPF}(a), instead of improving $Q(z^{-1})z^{-1}$ (Fig.~\ref{fig:LPF}(b)).
To correct the frequencies, $N$ is modified with
\begin{align}
	\label{eq:PDOB:Ncalculation}
	N=\frac{2\pi g\gamma-\omega_0}{T_{\mathrm{k}}g\omega_0\gamma}.
\end{align}
The derivation is as follows.
First, a correct delay $N$ is defined as
\begin{align}
	\label{eq:DEF:Nsigma}
	N&=\frac{2\pi}{T_{\mathrm{k}}\omega_0}(1+\sigma),
\end{align}
where $\sigma$ is a small variation.
Obtaining a fundamental frequency that causes the sensitivity function of zero, $1-e^{-j\omega_0T_{\mathrm{k}}}Q(e^{-j\omega_0 T_{\mathrm{k}}})=0$, is the objective, and it is rewritten by substituting the Q-filter in \eqref{eq:PDOB:Q} as
\begin{align}
	\frac{g\gamma(1-e^{-j\omega_0T_{\mathrm{k}}N})+j\omega_0}{g+j\omega_0}=0,\notag
\end{align}
where $z^{-1}$ is neglected.
The rewritten objective becomes
\begin{align}
	\label{eq:Eq:sigma}
	2\pi g\gamma\sigma+\omega_0 =0
\end{align}
with the approximation of the time delay element
\begin{align}
	e^{-j\omega_0T_{\mathrm{k}}N}
	=e^{-j2\pi}e^{-j2\pi\sigma}
	=e^{-j2\pi\sigma}\approx1-j2\pi\sigma\notag
\end{align}
based on \eqref{eq:DEF:Nsigma} and the small variation $\sigma$ satisfying $|2\pi\sigma|< 1$.
The larger $g$ is compared to $\omega_0$, the smaller is $\sigma$.
The modified delay calculation in \eqref{eq:PDOB:Ncalculation} is obtained by \eqref{eq:DEF:Nsigma} and \eqref{eq:Eq:sigma}.
\subsubsection{Design of $\gamma$}
The low-pass filter is set to $q(z^{-1})=1$ and the $Z$-operators $z^{-1}$ in $1-Q(z^{-1})z^{-1}$ and $Q(z^{-1})z^{-1}$ are neglected to simplify the design of $\gamma$.
The design parameter $\gamma$ especially affects $1-Q(z^{-1})$ and $Q(z^{-1})$ at the frequencies $\omega_{\mathrm{b1}}$ and $\omega_{\mathrm{b2}}$ with
\begin{align}
	\omega_{\mathrm{b1}}
	&=(2n)\frac{\omega_0}{2},\
	\omega_{\mathrm{b2}}
	=(2n+1)\frac{\omega_0}{2},\
	n=0,\ 1,\ 2,\ \ldots,\notag
\end{align}
as shown in Fig.~\ref{fig:gamma}.
The gains of the functions
\begin{align}
	|1-Q(e^{-j\omega T_{\mathrm{k}}})|
	&=\left|
	2\gamma\sin\left( -\frac{NT_{\mathrm{k}}}{2}\omega \right)
	\right|,\notag\\
	|Q(e^{-j\omega T_{\mathrm{k}}})|
	&=\left|
	\sqrt{1+ 4\gamma\left(\gamma-1\right) \sin^2 \left( -\frac{NT_{\mathrm{k}}}{2}\omega\right) }
	\right|\notag
\end{align}
become as follows for the frequencies $\omega_{\mathrm{b1}}$ and $\omega_{\mathrm{b2}}$
\begin{align}
	&|1-Q(e^{-j\omega_{\mathrm{b1}}T_{\mathrm{k}}})|=0,\
	|1-Q(e^{-j\omega_{\mathrm{b2}}T_{\mathrm{k}}})|=\left|2\gamma\right|,\notag\\
	&|Q(e^{-j\omega_{\mathrm{b1}}T_{\mathrm{k}}})|=1,\
	|Q(e^{-j\omega_{\mathrm{b2}}T_{\mathrm{k}}})|=\left|1 - 2\gamma\right|.\notag
\end{align}
Considering an optimal complementary sensitivity characteristic, $\gamma$ can be set to 0.5 to minimize $|Q(e^{-j\omega_{\mathrm{b2}}T_{\mathrm{k}}})|=\left|1 - 2\gamma\right|$.

\subsubsection{Design of $g$}
\begin{figure}[t]
		\begin{center}
			\includegraphics[width=0.85\hsize]{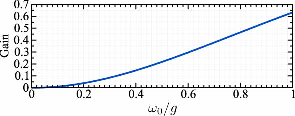}
		\end{center}
		\caption{Gain variation of \eqref{eq:omega_g} with respect to $\omega_0/g$ at the fundamental frequency.}
		\label{fig:Omega_g}
\end{figure}

The design parameter $\gamma$ is set to 0.5 and the $Z$-operators $z^{-1}$ in $1-Q(z^{-1})z^{-1}$ and $Q(z^{-1})z^{-1}$ are neglected to simplify the design of $g$.
The cutoff frequency $g$ of \eqref{eq:PDOB:qLPF} is designed in accordance with two objectives: periodic-disturbance suppression and robust stability.
First, a lower limit for $g$ is given by the fundamental-wave suppression performance.
At the fundamental frequency, the gain of $1-Q(e^{-j\omega_0T_{\mathrm{k}}})$ using \eqref{eq:PDOB:Ncalculation} is
\begin{align}
	\label{eq:omega_g}
	|1-Q(e^{-j\omega_0T_{\mathrm{k}}})|=
	\sqrt{\frac{\{1-\cos(2\mu)\} + 2\mu\{\mu-\sin(2\mu)\}}{2(1+\mu^2)}},
\end{align}
where $\mu=\omega_0/g$.
The gain depends only on $\omega_0/g$.
Fig.~\ref{fig:Omega_g} shows the gain variation with respect to $\omega_0/g$.
An upper limit for $\omega_0/g$ can be determined with the required precision and Fig.~\ref{fig:Omega_g}.
Since the fundamental frequency $\omega_0$ is given by a target periodic-disturbance, the lower limit for $g$ is obtained from the upper limit for $\omega_0/g$.

Next, an upper limit for $g$ is derived in accordance with the robust stability using the equivalent block diagram of the PDOB for the small-gain theorem shown in Fig.~\ref{fig:DOBs}(c).
The modeling error consists of the weighting function $W(z^{-1})$ and the variation $\delta(z^{-1})$
\begin{align}
	\Delta(z^{-1}) = W(z^{-1})\delta(z^{-1}),\notag
\end{align}
where the variation satisfies
\begin{align}
	\|\delta(z^{-1})\|_{\infty}\leq1.\notag
\end{align}
By assuming that the nominal PDOB and modeling error are stable, the robust-stability condition based on the small-gain theorem is
\begin{align}
	\|W(z^{-1})Q(z^{-1})z^{-1}\|_{\infty}<1.\notag
\end{align}
It can be rewritten as
\begin{align}
	\left|\frac{g}{g+j\omega}0.5(1+e^{-j\omega T_{\mathrm{k}}N})e^{-j\omega T_{\mathrm{k}}}\right|&\notag\\
	&\hspace{-20mm}\leq\left|\frac{g}{g+j\omega}\right|<\left|\frac{1}{W(e^{-j\omega T_{\mathrm{k}}})}\right|,\ \forall \omega.\notag
\end{align}
The upper limit for $g$ can be determined with this condition.
The simple expression $|\frac{g}{g+j\omega}|<|\frac{1}{W(e^{-j\omega T_{\mathrm{k}}})}|$ provides a sufficient condition without considering the time delay element $e^{-j\omega T_{\mathrm{k}}N}$.
\begin{figure}[t]
	\begin{center}
		\includegraphics[width=0.9\hsize]{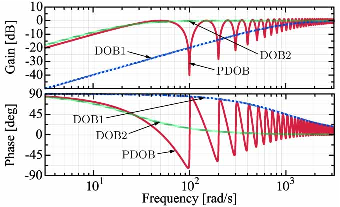}\\
		(a)\\
		\includegraphics[width=0.9\hsize]{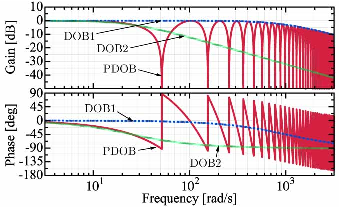}\\
		(b)
	\end{center}
	\caption{Comparison of PDOB with two types of DOB. The parameters for PDOB are \eqref{eq:PDOB:Ncalculation}, $\omega_0=100\ \mathrm{rad/s}$, $\gamma=0.5$, $g=1000\ \mathrm{rad/s}$, and $T_{\mathrm{k}}=0.01\ \mathrm{ms}$. The cutoff frequencies for DOB1 and DOB2 are $100\ \mathrm{rad/s}$ and $25\ \mathrm{rad/s}$, respectively. (a) Sensitivity functions, $1-Q(z^{-1})z^{-1}$. (b) Complementary sensitivity functions, $Q(z^{-1})z^{-1}$.}
	\label{fig:Comparison}
\end{figure}

\subsection{Comparison of PDOB with DOB} \label{sec2-3}
This subsection compares the PDOB with the DOB mentioned in \cite{DOBAna}.
The sensitivity function $1-Q(z^{-1})z^{-1}$ and the complementary sensitivity function $Q(z^{-1})$ are used in the comparison because the sensitivity characteristic corresponds to the disturbance suppression characteristic.
Moreover, the complementary sensitivity characteristic corresponds to the noise sensitivity and robust stability.
The Bode diagrams of the PDOB and two types of DOB are shown in Fig.~\ref{fig:Comparison}.
DOB1 and DOB2 use the cutoff frequencies 100 rad/s and 25 rad/s, respectively.
In the sensitivity functions shown in Fig.~\ref{fig:Comparison}(a), the PDOB achieves the lowest gain for the periodic disturbance, which consists of a fundamental wave at 100 rad/s and harmonics at 200, 300, … rad/s.
Compared with DOB2, the PDOB includes not only the high-pass characteristic but also an infinite number of band-stop characteristics.
Moreover, the complementary sensitivity function of the PDOB shown in Fig.~\ref{fig:Comparison}(b) has an infinite number of band-stop characteristics in addition to a low-pass characteristic in comparison with DOB1.
Therefore, both periodic-disturbance suppression characteristic and gain of the complementary sensitivity function of the PDOB are better than those of DOB1.
In comparison with DOB2, the PDOB improves the sensitivity function only at the frequencies of the periodic disturbance in the tradeoff.

\section{Adaptive Periodic-Disturbance Observer} \label{sec3}
\subsection{Fundamental Frequency Estimation} \label{sec3-1}
\begin{figure}[t]
	\begin{center}
			\includegraphics[width=0.9\hsize]{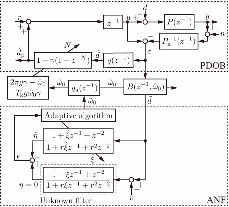}
	\end{center}
		\caption{Block diagram of adaptive PDOB.}
		\label{fig:AdaptiveMechanism}
\end{figure}

The PDOB requires the fundamental frequency $\omega_0$ of a periodic disturbance.
The suppression performance deteriorates when the fundamental frequency varies.
To correct the discrepancy, this paper also proposes an adaptive PDOB that recursively estimates the fundamental frequency.
A block diagram of the adaptive PDOB is shown in Fig.~\ref{fig:AdaptiveMechanism}.
In the ANF, the common cost function for a recursive-least-square algorithm \cite{AdaptiveFilterTheory} is used to estimate the correct $\xi(h)$ optimally
\begin{align}
	\label{eq:CostFunction}
	J(h)=\sum_{n=1}^{h}\lambda^{h-n}\{e(n)\}^2+\delta\lambda^h\{\hat{\xi}(h)\}^2,
\end{align}
where $\lambda$ and $\delta$ denote the forgetting factor and the regularization parameter, respectively.
The regularizing term $\delta\lambda^h\{\hat{\xi}(h)\}^2$ stabilizes the solution.
The algorithm of the ANF is derived in Appendix~\ref{app} and summarized in TABLE~\ref{tab:AdaptivePDOB}.
\begin{table}[t]
	\centering
	\caption{Algorithm of ANF.}\label{tab:AdaptivePDOB}
\begingroup
\renewcommand{\arraystretch}{1.0}
	\begin{tabular}{lc}
		\hline
		\hline
		\multicolumn{2}{l}{Parameters:}\\
		Identified fundamental freq. &$\hat{\omega}_0(0)$\\
		Output of the unknown filter &$\eta=0$\\
		Notch parameter 			 &$0<r<1$\\
		Multi-rate ratio 			 &$0<\kappa$\\
		Forgetting factor			 &$0\ll\lambda<1$\\
		Regularization parameter	 &$0<\delta$\\
		\hline
		\multicolumn{2}{l}{Initialization:}\\
		\multicolumn{2}{l}{$\hat{\xi}(0)=-2\cos{\{\hat{\omega}_0(0)T_{\mathrm{k}}\}},\
		P(0)=\delta^{-1}$}\\
		\hline
		\multicolumn{2}{l}{Notch filter:}\\
		\multicolumn{2}{l}{${\alpha}(k)=-r\hat{\eta}(k-1)+\tilde{d}(k-1)$}\\
		\multicolumn{2}{l}{${\beta}(k)=-r^2\hat{\eta}(k-2)+\tilde{d}(k)+\tilde{d}(k-2)$}\\
		\multicolumn{2}{l}{$\hat{\eta}(k)={\alpha}(k)\hat{\xi}(k)+{\beta}(k)$}\\
		\multicolumn{2}{l}{Adaptive algorithm:}\\
		\multicolumn{2}{l}{$g(h)=\{P(h-1){\alpha}(h)\}\{\lambda + P(h-1){\alpha}^2(h)\}^{-1}$}\\
		\multicolumn{2}{l}{$e(h)=\eta(h)-\hat{\eta}(h)$}\\
		\multicolumn{2}{l}{$\hat{\xi}(h)=\hat{\xi}(h-1)+g(h)e(h)$}\\
		\multicolumn{2}{l}{$P(h)={\lambda}^{-1}\{P(h-1)-g(h){\alpha}(h)P(h-1)\}$}\\
		\multicolumn{2}{l}{Fundamental frequency:}\\
		\multicolumn{2}{l}{$\tilde{\omega}_0(h)=T_{\mathrm{k}}^{-1}\cos^{-1}{\{-0.5\hat{\xi}(h)\}}$}\\
		\hline
		\hline
	\end{tabular}
\endgroup
\end{table}
It works as a frequency estimator for a pure sinusoidal wave.
The adaptive variable $\hat{\xi}$ that adapts to suppress the sinusoidal wave provides its frequency with $\tilde{\omega}_0(h)=T_{\mathrm{k}}^{-1}\cos^{-1}{\{-0.5\hat{\xi}(h)\}}$.

The adaptive PDOB has six additional design parameters: $r$, $\kappa$, $\lambda$, $\delta$, $g_{\mathrm{a}}$, and $g_{\mathrm{b}}$.
The parameters $r$ and $\kappa$ belong to the ANF; $\lambda$ and $\delta$ are defined in the cost function; $g_{\mathrm{a}}$ and $g_{\mathrm{b}}$ belong to the low-pass filter $q_{\mathrm{a}}(z^{-1})$ and the band-stop filter $B(z^{-1},\hat{\omega}_0)$, respectively.

\begin{figure}[t]
	\begin{center}
			\includegraphics[width=0.9\hsize]{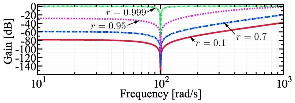}
	\end{center}
		\caption{Bode diagram of notch filter with variations in $r$. The other parameters are $\xi=-2\cos(100T_{\mathrm{k}})$ and $T_{\mathrm{k}}=0.1\ \mathrm{ms}$.}
		\label{fig:Notch}
\end{figure}

\begin{figure}[t]
	\begin{center}
			\includegraphics[width=0.9\hsize]{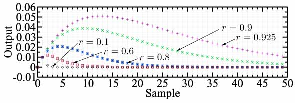}
	\end{center}
		\caption{Convergences of notch filter processing $\sin(100T_{\mathrm{k}}k)$ with variations in $r$. The other parameters are $\xi=-2\cos(100T_{\mathrm{k}})$ and $T_{\mathrm{k}}=0.1\ \mathrm{ms}$.}
		\label{fig:NotchConv}
\end{figure}

The notch parameter $r$ governs the bandwidth of the notch filter, shown in Fig.~\ref{fig:Notch}.
The multi-rate ratio $\kappa$ is for the calculation of the adaptive error $e=\eta-\hat{\eta}$, where the output of the unknown filter $\eta$ is set to zero because it is assumed to behave like an ideal notch filter.
However, as shown in Fig.~\ref{fig:NotchConv}, the ideal notch filter also needs a transient response to converge.
Thus, the adaptive algorithm is calculated under another sampling $h$, that is slower than $k$, to calculate $\eta-\hat{\eta}$ with steady-state outputs of the notch filters.
TABLE~\ref{tab:AdaptivePDOB} shows the sampling $h$ used by the adaptive algorithm.
An integer $\kappa$, that is a ratio of $T_{\mathrm{k}}$ and $T_{\mathrm{h}}$, is defined as a design parameter for the sampling $k$ and $h$ with
\begin{align}
	\kappa=T_{\mathrm{h}}/T_{\mathrm{k}},\notag
\end{align}
where $T_{\mathrm{k}}$ and $T_{\mathrm{h}}$ are the sampling times for $k$ and $h$, respectively.

\begin{figure}[t]
	\begin{center}
			\includegraphics[width=0.9\hsize]{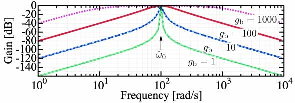}
	\end{center}
		\caption{Bode diagram of band-pass filter with variations in $g_{\mathrm{b}}$. The other parameters are $\hat{\omega}_0=100\ \mathrm{rad/s}$ and $T_{\mathrm{k}}=0.1\ \mathrm{ms}$.}
		\label{fig:BandPass}
\end{figure}

A band-pass filter $B(z^{-1},\hat{\omega}_0)$ is used to provide a pure fundamental wave as the input to the ANF $\tilde{d}$ from the output of the PDOB $\varepsilon$, which are shown in Fig.~\ref{fig:AdaptiveMechanism}.
This study uses the band-pass filter
\begin{align}
	&B(j\omega,\hat{\omega}_0)=\left\{\frac{j\omega g_{\mathrm{b}}}{(\hat{\omega}_0^2-\omega^2)+j\omega g_{\mathrm{b}}}\right\}^2,
\end{align}
which has a design frequency $g_{\mathrm{b}}$ governing the bandwidth, as show in Fig.~\ref{fig:BandPass}.
The parameter $g_{\mathrm{b}}$ needs to be small when harmonics of a periodic disturbance are strong.
The center frequency uses the fundamental frequency estimated by the ANF.
In addition to the band-pass filter, the low-pass filter $q_{\mathrm{a}}(z^{-1})$ is used to suppress oscillations in the outputted fundamental frequency $\tilde{\omega}_0$ due to other elements in $\tilde{d}$ that are not a fundamental wave.
The cutoff frequency $g_{\mathrm{a}}$ of the low-pass filter $q_{\mathrm{a}}(z^{-1})$ is also a design frequency.

\begin{figure}[t]
	\begin{center}
			\includegraphics[width=0.9\hsize]{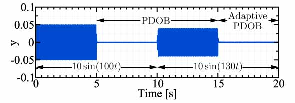}
	\end{center}
		\caption{Sinusoidal disturbance suppression using PDOB and adaptive PDOB. The parameters are $\hat{\omega}_0(0)=100\ \mathrm{rad/s}$, $\gamma=0.5$, $g=1000\ \mathrm{rad/s}$, $r=0.7$, $\kappa=10$, $\lambda=0.999$, $\delta=1000$, $g_{\mathrm{a}}=1000\ \mathrm{rad/s}$, $g_{\mathrm{b}}=1000\ \mathrm{rad/s}$, and $T_{\mathrm{k}}=0.1\ \mathrm{ms}$.}
		\label{fig:sim}
\end{figure}

A numerical example is shown in Fig.~\ref{fig:sim}.
It uses the following controller, plant, and disturbance:
\begin{align}
	r(z^{-1})&=-\left(200 + 10000\frac{T_{\mathrm{k}}}{1-z^{-1}}\right)y(z^{-1}), \notag\\
	y(z^{-1})&=\frac{T_{\mathrm{k}}}{1-z^{-1}}\left[\{r(z^{-1})+\hat{d}_{\mathrm{p}}(z^{-1})\}z^{-1}-d_{\mathrm{p}}(z^{-1})\right],\notag\\
	d_{\mathrm{p}}(k)& = \left\{
	\begin{array}{ll}
		\sin(100T_{\mathrm{k}}k)&\mathrm{if}\ T_{\mathrm{k}}k<10\ \mathrm{s}\\
		\sin(130T_{\mathrm{k}}k)&\mathrm{if}\ 10\ \mathrm{s}\leq T_{\mathrm{k}}k
	\end{array}
	\right. .\notag
\end{align}
The example validates the suppression characteristics of the PDOB and the adaptive PDOB.

\subsection{Convergence of Adaptive Algorithm} \label{sec3-2}
The convergence of the adaptive algorithm is described in this subsection.
A stationary environment that assumes a forgetting factor $\lambda$ of unity is considered.
In addition, $\tilde{d}(k)$ is assumed to be mainly composed of a fundamental wave.
Then, ${\alpha}(h)$ in the algorithm of the ANF mainly consists of a sinusoidal wave whose frequency is equal to that of the fundamental wave.
It guarantees that $\lim_{h\to\infty}1/\{\sum_{n=1}^{h}{\alpha}^2(n)\}$ converges to zero.
Based on the convergence, $P(h)$ in \eqref{eq:appendix:Covariance} also converges to zero with
\begin{align}
	\lim_{h\to\infty}P(h)&=\lim_{h\to\infty}\frac{1}{\sum_{n=1}^{h}{\alpha}^2(n)+\delta}=0.\notag
\end{align}
Consequently, the gain $g(h)$ in \eqref{eq:appendix:g} becomes zero.
Since $P(h)$ and $g(h)$ become zero in the steady-state, the steady-state $\hat{\xi}(h)$ in \eqref{eq:appendix:Xialg} is
\begin{align}
	\hat{\xi}(h)=\hat{\xi}(h-1).\notag
\end{align}
In conclusion, the convergences of $P(h)$, $g(h)$, and $\hat{\xi}(h)$ to zero and the steady-state are confirmed.

Next, the convergence of the adaptive variable $\hat{\xi}(h)$ to the true value $\xi(h)$ is described.
From \eqref{eq:appendix:Xi} and \eqref{eq:appendix:Covariance}, $\hat{\xi}(h)$ can be expressed by
\begin{align}
	\label{eq:Xi}
	\hat{\xi}(h)&=
	\frac{\sum_{n=1}^{h}{\alpha}(n)\{\eta(n)-{\beta}(n)\}}{\sum_{n=1}^{h}{\alpha}^2(n)+\delta}.
\end{align}
Because the output of the unknown filter is
\begin{align}
	\eta(n)&={\alpha}(n)\xi+\beta(n)-w(n),\notag
\end{align}
\eqref{eq:Xi} can be rewritten as
\begin{align}
	\hat{\xi}(h)&=
	\frac{\sum_{n=1}^{h}{\alpha}^2(n)}{\sum_{n=1}^{h}{\alpha}^2(n)+\delta}\xi
	-\frac{\sum_{n=1}^{h}{\alpha}(n)}{\sum_{n=1}^{h}{\alpha}^2(n)+\delta}w(n).\notag
\end{align}
The parameter $w(k)$ is noise effect due to $v(k)$ with
\begin{align}
	w(k)=&\{-rw(k-1)+v(k-1)\}{\xi}\notag\\
	&-r^2w(k-2)+v(k)+v(k-2),\notag
\end{align}
and $v(k)$ includes elements that are not a fundamental wave in $\tilde{d}(k)$, i.e., harmonics and aperiodic disturbances.
The noise effect $w(k)$ corresponds to the errors due to $\eta=0$.
In order to let $\hat{\xi}(h)$ converge to the true value $\xi(h)$, $\frac{\sum_{n=1}^{h}{\alpha}^2(n)}{\sum_{n=1}^{h}{\alpha}^2(n)+\delta}$ and $\frac{\sum_{n=1}^{h}{\alpha}(n)w(n)}{\sum_{n=1}^{h}{\alpha}^2(n)+\delta}$ need to be 1 and 0, respectively.
The regularization parameter $\delta$ adjusts them in the tradeoff: a small $\delta$ sets the first term to 1 and a large $\delta$ reduces the influence of $w(n)$.
Another way to improve the convergence is to adjust the band-pass filter $B(z^{-1},\hat{\omega}_0)$, which directly reduces the power of $w(n)$.

\begin{figure}[t]
	\centering
			\includegraphics[width=0.9\hsize]{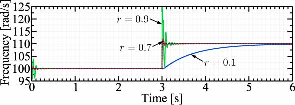}\\
			(a)\\
	\centering
			\includegraphics[width=0.9\hsize]{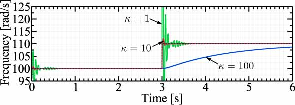}\\
			(b)\\
	\centering
			\includegraphics[width=0.9\hsize]{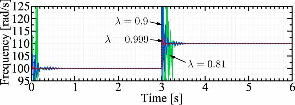}\\
			(c)\\
	\centering
			\includegraphics[width=0.9\hsize]{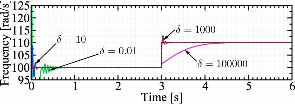}\\
			(d)\\
	\centering
			\includegraphics[width=0.9\hsize]{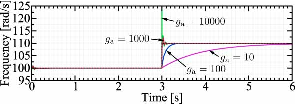}\\
			(e)\\
	\centering
			\includegraphics[width=0.9\hsize]{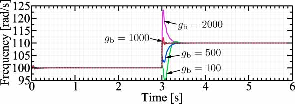}\\
			(f)\\
		\caption{Step frequency estimations for a sinusoidal wave. The standard parameters are $\hat{\omega}_0(0)=100\ \mathrm{rad/s}$, $r=0.7$, $\kappa=10$, $\lambda=0.999$, $\delta=1000$, $g_{\mathrm{a}}=1000\ \mathrm{rad/s}$, $g_{\mathrm{b}}=1000\ \mathrm{rad/s}$, and $T_{\mathrm{k}}=0.1\ \mathrm{ms}$.}
		\label{fig:SFE}
\end{figure}
\subsection{Step Frequency Estimation Examples for Design of Six Additional Parameters} \label{sec3-3}
\begin{figure}[t]
	\begin{center}
			\includegraphics[width=0.9\hsize]{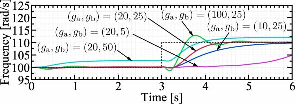}
	\end{center}
		\caption{Step frequency estimations for a periodic disturbance. The other parameters are $\hat{\omega}_0(0)=100\ \mathrm{rad/s}$, $r=0.7$, $\kappa=10$, $\lambda=0.999$, $\delta=1000$, and $T_{\mathrm{k}}=0.1\ \mathrm{ms}$.}
		\label{fig:harmonics}
\end{figure}

The design of the six additional design parameters $r$, $\kappa$, $\lambda$, $\delta$, $g_{\mathrm{a}}$, and $g_{\mathrm{b}}$ is described in this subsection with step frequency estimations verifying several parameter examples.
The estimation results are shown in Fig.~\ref{fig:SFE} in which the output of the PDOB, $\varepsilon$, is set to
\begin{align}
	\varepsilon(k) = \left\{
	\begin{array}{ll}
		\sin(100T_{\mathrm{k}}k)&T_{\mathrm{k}}k<3\ \mathrm{s}\\
		\sin(110T_{\mathrm{k}}k)&3\ \mathrm{s}\leq T_{\mathrm{k}}k
	\end{array}
	\right. .\notag
\end{align}
Fig.~\ref{fig:SFE}(a) shows that oscillations and overshoot occur when the notch parameter $r$ is large.
Hence, lower $r$ values are preferred.
After the determination of $r$, the multi-rate ratio $\kappa$ can be determined from $r$ and Fig.~\ref{fig:NotchConv} as a sufficiently large value to wait for the convergence of the notch filter.
Regarding the step frequency estimation shown in Fig.~\ref{fig:SFE}(b), a very small $\kappa$ that induces an oscillating response is not suitable.
The forgetting factor $\lambda$ is usually selected as a positive value close to, but less than, unity.
When set to a small value, $\lambda$ deteriorates the transient response, as shown in Fig.~\ref{fig:SFE}(c).
A design index of the regularization parameter $\delta$ is described in Section~\ref{sec3-2}.
A small $\delta$ realizes high-speed response and a large $\delta$ suppresses noise effect.
Moreover, Fig.~\ref{fig:SFE}(d) demonstrates that the initial response is much affected by $\delta$.
A large $\delta$ causes a smooth response.
In Figs.~\ref{fig:SFE}(e) and (f), the cutoff frequency $g_{\mathrm{a}}$ of the low-pass filter $q_{\mathrm{a}}(z^{-1})$ suppresses oscillations of the estimated fundamental frequency, and the design frequency $g_{\mathrm{b}}$ of the band-pass filter $B(z^{-1},\hat{\omega}_0)$ changes the transient response.
However, the two parameters should be designed in accordance with the example of the fundamental-frequency estimation from a periodic disturbance including also harmonics.
Fig.~\ref{fig:harmonics} shows step frequency estimations based on five combinations of $g_{\mathrm{a}}$ and $g_{\mathrm{b}}$.
The estimation uses an input including the fundamental frequency and ten harmonics:
\begin{align}
	\varepsilon(k) = \left\{
	\begin{array}{ll}
		\sum_{n=1}^{10}\sin(n100T_{\mathrm{k}}k)&T_{\mathrm{k}}k<3\ \mathrm{s}\\
		\sum_{n=1}^{10}\sin(n110T_{\mathrm{k}}k)&3\ \mathrm{s}\leq T_{\mathrm{k}}k
	\end{array}
	\right. .\notag
\end{align}
According to Fig.~\ref{fig:harmonics}, the cutoff frequency $g_{\mathrm{a}}$ suppresses the oscillations and the design frequency $g_{\mathrm{b}}$ modifies the steady-state accuracy.
The problems typically occur due to strong harmonics and aperiodic disturbances.
The frequencies $g_{\mathrm{a}}$ and $g_{\mathrm{b}}$ are determined with consideration of the characteristics and the convergence time.

\section{Simulations} \label{sec4}
\begin{table}[t!]
	\centering
	\caption{Simulation parameters.} \label{tab:SimPara}
		\begin{tabular}{lcl}
		\hline
		\hline
		Parameter & Symbol &Value\\
		\hline
		Sampling time					&$T_{\mathrm{k}}$				&0.1 ms\\
		Inertia							&${J}$				&$0.0028\ \mathrm{kgm^2}$\\
		Torque constant					&${K}_{\mathrm{t}}$			&$1.18\ \mathrm{Nm/A}$\\
		\hline
		Initial fundamental freq.		&${\omega}_{0}(0)$	&$10\ \mathrm{rad/s}$\\
		Design parameter				&${\gamma}$		&0.7\\
		Cutoff freq. of $q(z^{-1})$ 	&${g}$ 			&1000 rad/s\\
		Notch parameter					&${r}$			&0.7\\
		Multi-rate ratio				&${\kappa}$		&10\\
		Forgetting factor				&${\lambda}$	&0.999\\
		Regularization parameter		&${\delta}$		&1000\\
		Cutoff freq. of $q_{\mathrm{a}}(z^{-1})$ &${g}_{\mathrm{a}}$ 	&1 rad/s\\
		Design freq. of $B(z^{-1})$				 &${g}_{\mathrm{b}}$ 	&1 rad/s\\
		\hline
		Cutoff freq. for the DOB and RC	&${g}$ 			&1000 rad/s\\
		\hline
		\hline
		\end{tabular}
\end{table}
Two simulations to compare the PDOB with an RC and DOB and the PDOB with the adaptive PDOB were conducted.
These controller, plant, and disturbance were assumed
\begin{align}
	I^{\mathrm{ref}}(z^{-1})=&-\frac{J}{K_{\mathrm{t}}}\left(2500 + 100D(z^{-1})\right)x^{\mathrm{res}}, \notag\\
	x^{\mathrm{res}}(z^{-1})=&\frac{1}{J}\left(\frac{T_{\mathrm{k}}}{1-z^{-1}}\right)^2\notag\\
	&\left[\{K_{\mathrm{t}}I^{\mathrm{ref}}(z^{-1})+\hat{d}_{\mathrm{p}}(z^{-1})\}z^{-1}-d_{\mathrm{p}}(z^{-1})\right],\notag\\
	d_{\mathrm{p}}=&\sum_{n=1}^{20}\sin(n\omega_0(k)T_{\mathrm{k}}k),\notag
\end{align}
where $D(z^{-1})$ is the pseudo differentiator.
The parameters are shown in TABLE~\ref{tab:SimPara}.
The modified RC with $a=1$ \cite{RC1} and DOB \cite{DOBAna} were used.
The command was set to zero to validate only the disturbance suppression characteristics.
The fundamental frequency $\omega_0(k)$ was constantly set to 10 rad/s in Simulation 1.
Simulation 2 used
\begin{align}
	\omega_0(k)&=
	\left\{
	\begin{array}{ll}
			10\ \mathrm{rad/s}&\mathrm{if}\ 0\ \mathrm{s}\leq T_{\mathrm{k}}k < 40\ \mathrm{s}\\
			11\ \mathrm{rad/s}&\mathrm{if}\ 40\ \mathrm{s}\leq T_{\mathrm{k}}k\\
	\end{array}
	\right. .\notag
\end{align}

\begin{figure}[t!]
	\begin{center}
		\includegraphics[width=0.9\hsize]{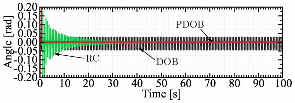}
	\end{center}
		\caption{Results of Simulation 1.}
		\label{fig:Sim1}
\end{figure}

\begin{figure}[t!]
	\begin{center}
		\includegraphics[width=0.9\hsize]{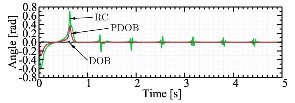}
	\end{center}
		\caption{Transient responses of Fig.~\ref{fig:Sim1}.}
		\label{fig:Sim1Tr}
\end{figure}

\begin{figure}[t!]
	\begin{center}
		\includegraphics[width=0.9\hsize]{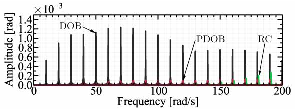}
	\end{center}
		\caption{DFT results of Fig.~\ref{fig:Sim1} for 20 s - 100 s.}
		\label{fig:Sim1fft}
\end{figure}

The results of Simulation 1 are shown in Fig.~\ref{fig:Sim1} and the transient responses of Fig.~\ref{fig:Sim1} are illustrated in Fig.~\ref{fig:Sim1Tr}.
The steady-state characteristics are presented by using a discrete Fourier transform (DFT).
The DFT results of Fig.~\ref{fig:Sim1} from 20 s - 100 s are shown in Fig.~\ref{fig:Sim1fft}.
In the transient response, the RC and PDOB need a storing process to be efficient due to their time delay elements.
However, the PDOB achieves a better transient response than the RC.
In the steady-state between 20 s - 100 s, the PDOB shows the better suppression performance at all frequencies of the periodic disturbance than that of the DOB.

\begin{figure}[t!]
	\begin{center}
		\includegraphics[width=0.9\hsize]{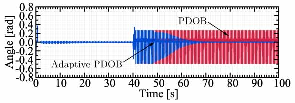}
	\end{center}
		\caption{Results of Simulation 2.}
		\label{fig:Sim2}
\end{figure}

\begin{figure}[t!]
	\begin{center}
		\includegraphics[width=0.9\hsize]{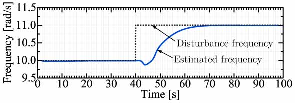}
	\end{center}
		\caption{Fundamental-frequency estimation result of adaptive PDOB.}
		\label{fig:Sim2freq}
\end{figure}

The results of Simulation 2 are shown in Fig.~\ref{fig:Sim2}.
The fundamental-frequency estimation results of the adaptive PDOB are depicted in Fig.~\ref{fig:Sim2freq}.
The transient response of the adaptive estimation needs a long convergence time due to the harmonics between 40 s - 70 s.
However, the adaptivity, which together with the estimation of the varying frequency achieves the best suppression performance, can be confirmed.

\section{Experiments} \label{sec5}
\begin{table}[t!]
	\centering
	\caption{Experimental parameters.} \label{tab:ExpPara}
		\begin{tabular}{lcl}
		\hline
		\hline
		Parameter & Symbol &Value (1st, 2nd, 3rd) Joint\\
		\hline
		Sampling time		&$T_{\mathrm{k}}$					&0.1 ms\\
		Proportional gain	&$\bm{K}_{\mathrm{P}}$				&$(400,\ 400,\ 400)$\\
		Differential gain	&$\bm{K}_{\mathrm{D}}$				&$(40,\ 40,\ 40)$\\
		Nominal inertia				&$\bm{J}_{\mathrm{n}}$				&$(7.0,\ 3.0 ,\ 0.3)\ \mathrm{kgm^2}$\\
		Nominal torque constant		&$\bm{K}_{\mathrm{tn}}$				&$(0.59,\ 0.59,\ 0.238)\ \mathrm{Nm/A}$\\
		Gear ratio			&$\bm{G}_{\mathrm{r}}$				&$(192,\ 120,\ 80)$\\
		Length				&$\bm{L}$							&$(0.26,\ 0.27,\ 0.09)\ \mathrm{m}$\\
		\hline
		Identified fundamental freq.	&$\bm{\omega}_{0}(0)$	&$(3,\ 3,\ 3)\ \mathrm{rad/s}$\\
		Design parameter				&$\bm{\gamma}$		&$(0.5,\ 0.5,\ 0.5)$\\
		Cutoff freq. of $q(z^{-1})$ 	&$\bm{g}$ 			&$(200,\ 200,\ 200)$ rad/s\\
		Notch parameter					&$\bm{r}$			&$(0.1,\ 0.1,\ 0.1)$\\
		Multi-rate ratio				&$\bm{\kappa}$		&$(10,\ 10,\ 10)$\\
		Forgetting factor				&$\bm{\lambda}$		&$(0.999,\ 0.999,\ 0.999)$\\
		Regularization parameter		&$\bm{\delta}$		&$(10^7,\ 10^7,\ 10^7)$\\
		Cutoff freq. of $q_{\mathrm{a}}(z^{-1})$ &$\bm{g}_{\mathrm{a}}$ 	&$(1,\ 1,\ 1)$ rad/s\\
		Design freq. of $B(z^{-1})$				 &$\bm{g}_{\mathrm{b}}$ 	&$(2,\ 2,\ 2)$ rad/s\\
		\hline
		Cutoff freq. for the DOB 		&$\bm{g}$ 			&$(200,\ 200,\ 200)$ rad/s\\
		Cutoff freq. for the RC 		&$\bm{g}$ 			&$(30,\ 30,\ 30)$ rad/s\\
		\hline
		\hline
		\end{tabular}
\end{table}
\begin{table}[t]
	\centering
	\caption{RMSEs of experimental results.} \label{tab:ExpRMSE}
		\begin{tabular}{llll}
		\hline
		\hline
		Experiment 1 & DOB & RC + DOB & PDOB + DOB\\
		$x$-axis	&0.256 mm&0.157 mm&0.066 mm\\
		$y$-axis	&0.277 mm&0.150 mm&0.073 mm\\
		\hline
		Experiment 2 & PDOB + DOB & \multicolumn{2}{l}{Adaptive PDOB + DOB}\\
		$x$-axis	&0.191 mm&\multicolumn{2}{l}{0.117 mm}\\
		$y$-axis	&0.236 mm&\multicolumn{2}{l}{0.127 mm}\\
		\hline
		\hline
		\end{tabular}
\end{table}
\begin{figure}[t!]
	\begin{center}
			\includegraphics[width=0.6\hsize]{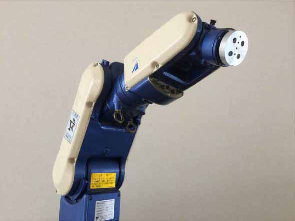}
	\end{center}
		\caption{Experimental multi-axis manipulator.}
		\label{fig:ExpSys}
\end{figure}
\begin{figure}[t]
	\begin{center}
			\includegraphics[width=0.9\hsize]{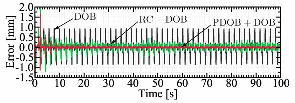}
	\end{center}
		\caption{Error values regarding $y$-axis of Experiment 1.}
		\label{fig:Exp1}
\end{figure}
\begin{figure}[t!]
	\begin{center}
			\includegraphics[width=0.9\hsize]{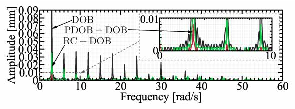}
	\end{center}
		\caption{DFT results of Fig.~\ref{fig:Exp1} from 20 s - 100 s.}
		\label{fig:Exp1fft}
\end{figure}
\begin{figure}[t!]
	\begin{center}
			\includegraphics[width=0.9\hsize]{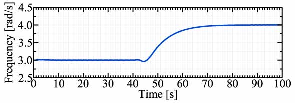}
	\end{center}
		\caption{Fundamental-frequency estimation result of adaptive PDOB implemented at first joint.}
		\label{fig:Exp2F}
\end{figure}
\begin{figure}[t!]
	\begin{center}
			\includegraphics[width=0.9\hsize]{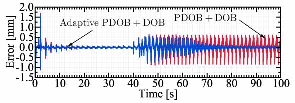}
	\end{center}
		\caption{Error values regarding $y$-axis of Experiment 2.}
		\label{fig:Exp2}
\end{figure}
Two experiments to compare the PDOB with an RC and DOB and the PDOB with the adaptive PDOB were conducted with the multi-axis manipulator shown in Fig.~\ref{fig:ExpSys}.
The parameters are summarized in TABLE~\ref{tab:ExpPara} and the commands in the work space were given as
\begin{align}
	x^{\mathrm{cmd}}(k)&=x_{0} + 50\cos\{\omega_0(k)T_{\mathrm{k}}k\}-50\ \mathrm{mm}\notag\\
	y^{\mathrm{cmd}}(k)&=y_{0} + 50\sin\{\omega_0(k)T_{\mathrm{k}}k\}\ \mathrm{mm},\notag
\end{align}
where $x_0$ and $y_0$ are the initial positions.
In Experiment 1, the command frequency $\omega_0(k)$ was constantly set to $3\ \mathrm{rad/s}$.
In Experiment 2, it was set to
\begin{align}
	\omega_0(k)&=
	\left\{
	\begin{array}{ll}
			3\ \mathrm{rad/s}&\mathrm{if}\ 0\ \mathrm{s}\leq T_{\mathrm{k}}k < 40\ \mathrm{s}\\
			4\ \mathrm{rad/s}&\mathrm{if}\ 45\ \mathrm{s}\leq T_{\mathrm{k}}k\\
	\end{array}
	\right. .\notag
\end{align}
In the shifting phase of $40\ \mathrm{s}$ - $45\ \mathrm{s}$, a smooth command that shifts $x^{\mathrm{cmd}}_1$ to $x^{\mathrm{cmd}}_2$ was given as
\begin{align}
	x^{\mathrm{cmd}}=x^{\mathrm{cmd}}_1+(x^{\mathrm{cmd}}_2-x^{\mathrm{cmd}}_1)(T_{\mathrm{k}}k-40)/(45-40).\notag
\end{align}
In the experimental control systems, the modified RC with $a=1$ \cite{RC1} and DOB \cite{DOBAna} were applied.
All control methods were implemented with a proportional-derivative controller and a feedforward controller that sets the input-output transfer function to 1.
Hence, the experimental errors depend only on disturbance suppression characteristics.
More specifically, the RC, PDOB, and adaptive PDOB were implemented with the DOB because they do not compensate aperiodic disturbances.

The root-mean-square errors (RMSEs) of the experimental results are shown in TABLE~\ref{tab:ExpRMSE}.
The error values regarding the $y$-axis of Experiment 1 are shown in Fig.~\ref{fig:Exp1}.
In the transient response, the RC and PDOB require storing time to be efficient because of their time delay elements.
In the steady-state, the PDOB and DOB combination performs the best periodic-disturbance suppression.
The DFT results of Fig.~\ref{fig:Exp1} are shown in Fig.~\ref{fig:Exp1fft}.
They demonstrate that the PDOB with DOB achieve suppression of all fundamental wave and harmonics in the frequency domain.
Further, the PDOB does not deteriorate the compensation by the DOB, whereas the RC amplifies the other disturbances for frequencies between those of the periodic disturbance in the frequency domain larger than 20 rad/s.
In Experiment 2, the compensation of the frequency-varying periodic disturbance was validated.
The adaptivity estimated the fundamental frequency of the periodic disturbance, as shown in Fig.~\ref{fig:Exp2F}, and maintains the best suppression performance of the PDOB.
The error values regarding the $y$-axis are shown in Fig.~\ref{fig:Exp2}.
In the steady-state, the adaptive PDOB achieves the best performance of the PDOB.
However, the adaptive estimation and the performance improvement need a long convergence time.

\section{Conclusion} \label{sec5}
The PDOB and adaptive PDOB were proposed in this paper.
The PDOB was constructed to compensate a periodic disturbance with three parameters: a fundamental frequency of a periodic disturbance $\omega_0$, design parameter $\gamma$, and cutoff frequency $g$.
Because the fundamental frequency $\omega_0$ needs preidentification, the adaptive PDOB including an ANF was also proposed with six additional design parameters: a notch parameter $r$, multi-rate ratio $\kappa$, forgetting factor $\lambda$, regularization parameter $\delta$, cutoff frequency $g_{\mathrm{a}}$, and design frequency $g_{\mathrm{b}}$.
The adaptivity causes the PDOB to achieve the best performance and realizes a compensation of frequency-varying periodic disturbances by estimating the fundamental frequency of a periodic disturbance.

\appendices
\section{Derivation of Adaptive Algorithm} \label{app}
This appendix presents the derivation of the ANF algorithm.
The ANF is expressed with $\hat{\eta}(k)=\alpha(k)\hat{\xi}(k)+\beta(k)$, as shown in TABLE~\ref{tab:AdaptivePDOB}.
The algorithm for the adaptive variable $\hat{\xi}(h)$ is obtained in accordance with the minimization of the cost function in \eqref{eq:CostFunction} with respect to $\hat{\xi}(h)$.
The condition $\frac{\partial J(h)}{\partial \hat{\xi}(h)}=0$, which satisfies the minimization of the cost function, provides
\begin{align}
	2\sum_{n=1}^h\lambda^{h-n}{\alpha}(n)\{\eta(n)-{\alpha}(n)\hat{\xi}(h)-{\beta}(n)\}-2\delta\lambda^h\hat{\xi}(h)=0.\notag
\end{align}
The $\hat{\xi}(h)$ satisfying $\frac{\partial J(h)}{\partial \hat{\xi}(h)}=0$ is obtained with
\begin{align}
	\label{eq:appendix:Xi}
	\hat{\xi}(h)&=P(h)\sum_{n=1}^{h}\lambda^{h-n}{\alpha}(n)\{\eta(n)-{\beta}(n)\},
\end{align}
where $P(h)$ is
\begin{align}
	\label{eq:appendix:Covariance}
	P(h)&=\frac{1}{\sum_{n=1}^{h}\lambda^{h-n}{\alpha}^2(n)+\delta\lambda^h}.
\end{align}
From \eqref{eq:appendix:Covariance}, the recursive $P^{-1}(h)$ is derived as
\begin{align}
	\label{eq:appendix:CovarianceInv}
	P^{-1}(h)&=\lambda P^{-1}(h-1) + {\alpha}^2(h).
\end{align}
Further, \eqref{eq:appendix:Xi} is transformed into
\begin{align}
	\hat{\xi}(h)=\hat{\xi}(h-1)+P(h){\alpha}(h)e(h).\notag
\end{align}
The calculation of $\hat{\xi}(h)$ for the algorithm is obtained as
\begin{align}
	\label{eq:appendix:Xialg}
	\hat{\xi}(h)=\hat{\xi}(h-1)+g(h)e(h),
\end{align}
where the gain $g(h)$ is
\begin{align}
	\label{eq:appendix:g}
	g(h)=P(h){\alpha}(h).
\end{align}
The adaptive error is calculated with $e(h)=\eta(h)-\hat{\eta}(h)$, and the gain is rewritten by substituting \eqref{eq:appendix:CovarianceInv} with \eqref{eq:appendix:g} for a recursive equation
\begin{align}
	\label{eq:appendix:galg}
	g(h)=\frac{P(h-1){\alpha}(h)}{\lambda + P(h-1){\alpha}^2(h)}.
\end{align}
$P(h)$ is further rewritten with \eqref{eq:appendix:CovarianceInv} and \eqref{eq:appendix:galg} as
\begin{align}
	P(h)=\frac{1}{\lambda}\{P(h-1)-g(h){\alpha}(h)P(h-1)\}.\notag
\end{align}

\section*{Acknowledgment}
This work was partially supported by JSPS KAKENHI.


\newpage
\begin{IEEEbiography}[{\includegraphics[width=\hsize]{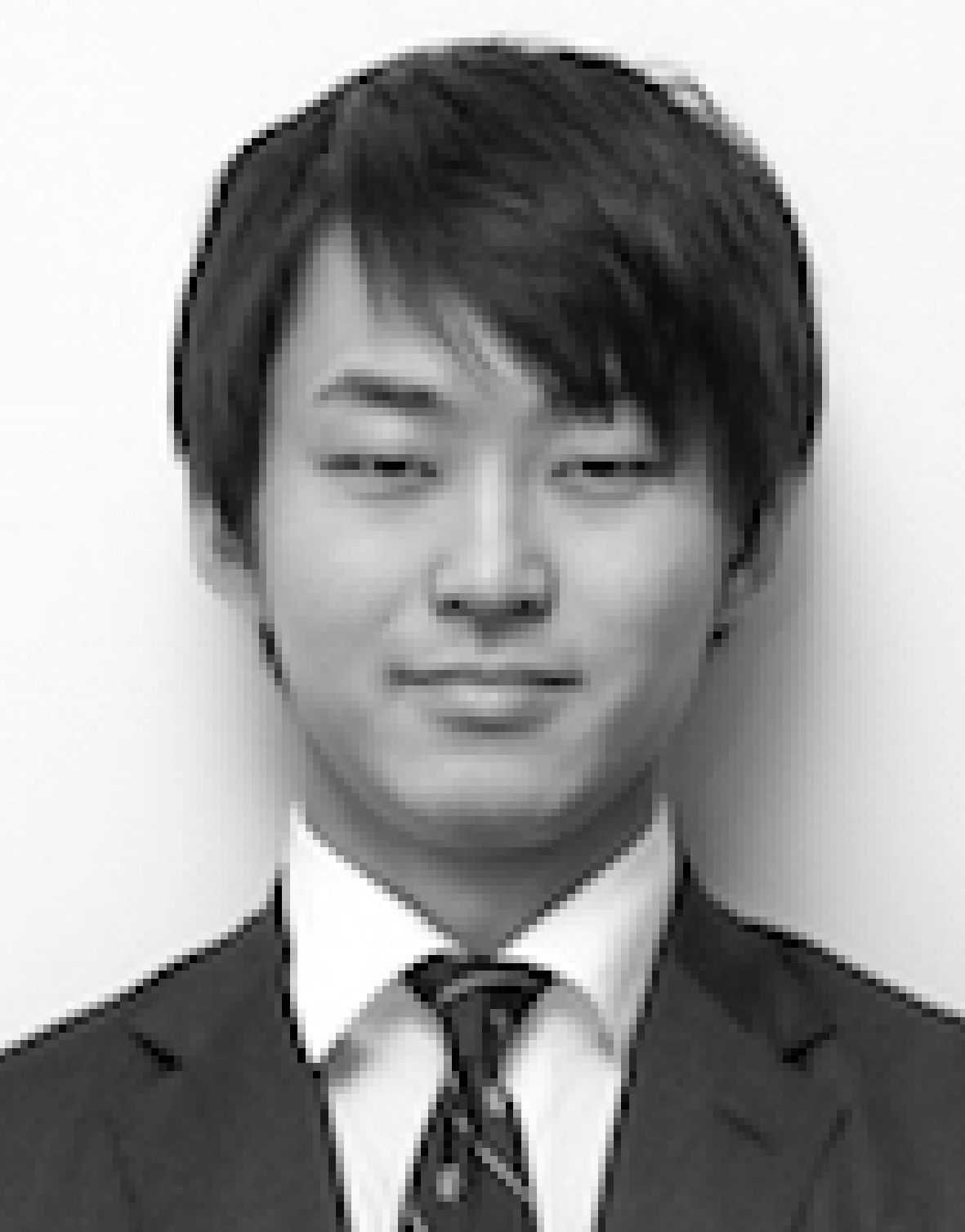}}]{Hisayoshi Muramatsu (S'16)}
received the B.E. degree in system design engineering and the M.E. degrees in integrated design engineering from Keio University, Yokohama, Japan, in 2016 and 2017, respectively.
He is currently working toward the Ph.D. degree in integrated design engineering from Keio University, Yokohama, Japan.
His research interests include motion control and mechatronics.
\end{IEEEbiography}
\vspace{-150mm}
\begin{IEEEbiography}[{\includegraphics[width=\hsize]{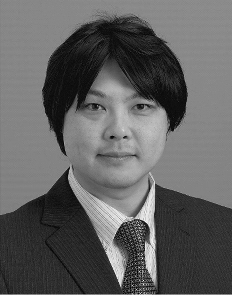}}]{Seiichiro Katsura (S'03-M'04)}
received the B.E. degree in system design engineering and the M.E. and Ph.D. degrees in integrated design engineering from Keio University, Yokohama, Japan, in 2001, 2002 and 2004, respectively.

From 2003 to 2005, he was a Research Fellow of the Japan Society for the Promotion of Science (JSPS).  From 2005 to 2008, he worked at Nagaoka University of Technology, Nagaoka, Niigata, Japan.  Since 2008, he has been at Keio University, Yokohama, Japan.  In 2017, he was a Visiting Researcher with The Laboratory for Machine Tools and Production Engineering (WZL) of RWTH Aachen University, Aachen, Germany.  His research interests include applied abstraction, human support, wave system, systems energy conversion, and electromechanical integration systems.

Prof. Katsura serves as an Associate Editor of the IEEE Transactions on Industrial Electronics. He was the recipient of The Institute of Electrical Engineers of Japan (IEEJ) Distinguished Paper Awards in 2003 and 2017, IEEE Industrial Electronics Society Best Conference Paper Award in 2012, and JSPS Prize in 2016.
\end{IEEEbiography}

\end{document}